# Coarse-grained periodic orbits and bifurcations in a Markov chain model for evolution of labor division

Lei Chai, Dahui Wang, Jiawei Chen and Zengru Di*

*Department of Systems Science, School of Management, Beijing Normal University, Xinjiekouwai street 19, Beijing, 100875, P.R. China*

## Abstract

We construct a Markov process model to describe the evolution of labor division and its dynamical behavior is investigated by numerical simulations in detail. We have shown that under the mechanism of increasing returns, the division of labor will emerge from the initial homogenous distribution and with the change of parameters there are period-adding coarse-grained bifurcations in the model. The model gives an example of emergent property in the evolution of complex systems and reveals an interesting dynamical behavior of the Markov chain process.

PACS number(s): 87.23.-n, 89.75.Fb, 05.45.-a

## I. INTRODUCTION

The division of labor leading to task specialization and greater colony efficiency is thought to be the major benefit of insect eusociality[1, 2]. Yet the mechanics of labor division is still an intricate puzzle and the exploring of it has become a major subject in many fields including both nature and social sciences during last decades [3, 4]. In order to understand how social organization evolves, we should understand the mechanisms that link the different levels among them [5]. In the context of physics and chemistry, theories of self-organization have been developed to describe the emergence of macroscopic patterns rising from interactions confined in microscopic level. The theories can be extended to biology and economy to show that complex collective behaviors may emerge from interaction among individuals that exhibit simple behaviors [6]. Actually, nonlinear dynamics has been widely applied to the analysis of biological problems, from traditional subject of population dynamics[7, 8] to the pattern formation and complex behavior within the process of evolution[9, 10].

---

* Electronic address: zdi@bnu.edu.cn

The recent developments in the study of collective behaviors of multi-agent system also provide us illuminating visions [11]. Although the interactions are simple and local, they may lead to complex dynamics at the global scale.

The division of labor can also be simulated in such systems and we have previously set up a model [12] to investigate the formation of it. Supposing a system is composed of many individual agents that have no complete information about the environment. There are two kinds of tasks for every agent: searching for unknown resources and exploiting the resource that they have found. The behavior of every agent is characterized by the probability for it to look for or to exploit resources. The charactor can be changed in the process of evolution. A Markov chain model is used to study the optimal evolutionary process of the system. In this paper, we will dig emphatically on the dynamical of mechanisms undergoing the division of labor in the system. As we shall see, with the selection acting on agent specialization at the level of system and under the condition of increasing returns, the division of labor emerges as the results of long-term optimizing evolution. The Markov chain model has shown some novel and interesting bifurcation phenomena, including coarse-grained periodic orbits and bifurcations. These results provide insights useful to understand the evolutionary dynamics and the formation of organization in biology and economy. And they are also helpful for us to get deep understanding of nonlinear dynamical systems.

This paper is organized as following. In the next section a Markov chain model has been set up and described. For the purpose of global-optimization, the best distribution of the labor is provided by the evolutionary dynamics of the system. We find that when the mechanism of increasing returns exceed a critical value, the labor distribution in the system will finally achieves a specialization as the results of long term optimum evolution. In section III, we analyze graphs depicting the phase space of the above dynamical system, in which attractors, bifurcations and chaos are included. In section IV, we provide a summary and a brief discussion of our results.

## II. MARKOV CHAIN MODEL FOR THE OPTIMAL EVOLUTION

We consider a colony made up of multi-agents trying to gain optimum benefit from the environment containing resources. The agents have no complete information about the resources. Each agent has two behaviors to choose: searching for unknown resources and exploiting resources already known. The agents' behaviors, which can be changed in the process of evolution from generation to generation, are characterized by real numbers depicting the probability for searching or exploiting. We apply Markov chain model to study the optimal evolution process of this system. As it will be discussed later, under the mechanism of increasing return and when its effect exceeds a critical value, the system will gain certain macroscopic structures by long-term evolution.

Now consider a system made up of $M$ agents, living in an environment containing resources valued $F_0$ and 0. Each agent is characterized by a parameter $q_i$ ($0 \le q_i \le 1, i=0,1,2,\ldots M$) determines the searching probability of it. At the beginning of every period, agents can choose to search new resources with probability $q$ or to exploit resources with probability 1-$q$. And if one decides to search, there is a probability $P$ for it to reach the resource $F_0$. Once one agent has found a $F_0$, all the others would go and exploit it. The benefit of agent $i$ is $(1+a(1-q_i))F_0$. When the agents have not found the resource $F_0$, they stay in the last position. Then the value of resource is reduced due to the exploiting of the agents. Hence if there is no new resource, the value of the last resource will be changed to

$$F_t = \frac{1}{a} F_{t-1}. \tag{1}$$

There is also a searching cost $a(1- q_i)C$ for any agent. In the above two formulas of benefit and cost, factor $a(1- q_i)$ is related with the intensity of the mechanism of increasing returns, and by changing the value of parameter $\alpha$ we can explore the effects of increasing return on the system's evolution.

For the system introduced above, assuming that there are $k$+1 states corresponding to the searching probabilities of the agents described by parameter $q_j$, among which there is $q_0$=0 and $q_k$=1. We use the distribution of agents, i.e., the number of agents in every state, to describe the state of the system in macro-level. We denote this distribution as $\{N_j, j=0, 1, \cdots\cdots, k\}$, and $\sum_{j=0}^{k} N_j = M$. For simplicity and without losing generality, in the following discussion we assume that $N_j$ is a positive real number instead of a positive integer. Then for any given distribution of agents, there is equivalently $m = \sum_{j=0}^{k} N_j q_j$ agents search for resources and the probability for them to find one resource $F_0$ is $P_r = 1-(1-P)^m$. Hence the expected returns of every agent in each period are:

$$E_0 = P_r((1+\alpha)M - \alpha m)F_0,$$

$$E_1 = E_0 + (1-P_r)\frac{1}{a}E_0,$$

$$E_2 = E_0 + (1-P_r)\frac{1}{a}E_0 + (1-P_r)^2\frac{1}{a^2}E_0,$$

……

Let $x=(1-P_r)/a$, where parameter $a$>1 is related to the diminishing of resource under $M$ agents' exploiting, the benefit of the $n$th period can be written as

$$E_n = E_0\left(\frac{1-x^n}{1-x}\right). \tag{2}$$

Hence the total benefit of one generation with $T$ periods is

$$
\begin{aligned}
W &= \sum_{n=0}^{T-1} E_0 \left(\frac{1-x^n}{1-x}\right) = \frac{E_0}{1-x}\left(T - x\left(\frac{1-x^T}{1-x}\right)\right) \\
&= \frac{P_r((1+\alpha)M - \alpha m)F_0}{1-x}\left(T - x\left(\frac{1-x^T}{1-x}\right)\right)
\end{aligned}
\qquad (3)
$$

The total cost $C_D$ of the system in one generation with $T$ periods is

$$
C_D = T\sum_{j=1}^{k} N_j q_j (1+\alpha(1-q_j))C . \qquad (4)
$$

Then the total return of one generation is

$$
R(\{N_j\}) = W(\{N_j\}) - C_D(\{N_j\}) . \qquad (5)
$$

Figure1

In the long-term evolution, there is a variation process between two generations. The state of every agent would transit among all the $k$+1 states. We suppose there is only transition between the nearest neighbors. We introduce a Markov chain process as shown in Figure 1 to describe the genetic variation and natural selection in the evolution. The dynamical behavior of this Markov chain is determined by following equations:

$$
\begin{aligned}
N_j(t+1) = N_j(t) + P(j+1 \to j)N_{j+1}(t) + P(j-1 \to j)N_{j-1}(t) \\
- P(j \to j-1)N_j(t) - P(j \to j+1)N_j(t)
\end{aligned}
\qquad (6)
$$

For $j$=0 and $j$=$k$ we have

$$
N_0(t+1) = N_0(t) + P(1 \to 0)N_1(t) - P(0 \to 1)N_0(t) ,
$$

$$
N_k(t+1) = N_k(t) + P(k-1 \to k)N_{k-1}(t) - P(k \to k-1)N_k(t) . \qquad (7)
$$

In the equation (6) and (7), $P(i \to j)$ is the transition probability for an agent from state $i$ to state $j$. It is determined by the global optimization. Corresponding to the natural selection process that is copy or replacement of a varying agent according to its result on total returns, this transition probability could be writing as

$$
P(i \to j) = \frac{1}{2}\mu[5 + 3\,\mathrm{sgn}(R(N_i - 1, N_j + 1) - R(N_i, N_j))] , \qquad (8)
$$

where $\mu$ is a parameter related to the probability of variation for every agent. From the above dynamical system, we can get the results of evolution of the system from any given initial conditions.

## III. NUMERICAL RESULTS OF THE DYNAMICAL PROPERTIES

In this section, we focus on the dynamical behavior of the system to investigate the structure of phase space with bifurcations and attractors. The numerical examples of this section are obtained with particular sets of parameter, and the phenomena that we describe below are persistent and could also be observed in other several parameter regimes. We consider a system with $M$=50 agents. The other parameters are: $F_0$ =10.0, $C$=2.0, $P$=0.4, $a$=5.0, $T$=10.0, $\mu$=0.04. The number of states $k$=11, that means the $i$th agent's character could adopt following values: $q_j$=0, 0.1, 0.2,…, 0.9, 1.0. In the initial the agents are homogeneous. And we set every agent has the same behavior character $q_j$=0.5. Given a certain value for parameter $\alpha$, that describes the intensity of increasing returns, we can get the evolution behavior of the system by the Markov chain model (3)-(8).

A. Evolution and bifurcations

Given any initial distribution, the system will evolve according to the mechanism we have developed above. We are especially interested in the asymptotic behavior of the system. For the sake of simplicity, we take one parameter $\alpha$ under variation. Figure 2 shows the evolving outcomes. They are different with several of the value of parameters. It has been found from the simulations that only when the parameter $\alpha$ exceed a certain value, could the system achieves the fully specialization as the result of long-term optimizing evolution.

Figure 2

In Figure 2 (a), where $\alpha$ = 0.04, the agents in the system have not been divided into two groups in the final. When $\alpha$=0.15, as shown in Figure 2 (b), the distribution of agents obviously aggregates in to two groups near $q_i$=0 and $q_i$=1. In order to explore the effect of changing parameters, we focus on the state where $q_i$=0 with the sequent change of $\alpha$ while other parameters remain constant. As the increasing of $\alpha$, the number of agents at the state $q_i$=0 reveals intricate bifurcations and periodic transformations as Figure 3 shows. Figure 4 is the corresponding change of total returns. Although it increases proportional to $\alpha$, we can see its detailed structures by a certain kind of normalization. We could find that it undergoes correspondent change as Figure 3. Because the above long-term results are all from the same initial distribution, even though they are related with the bifurcations of the system, the above figures are not exactly the bifurcation diagrams. We will investigate them in next subsection in detail.

Figure 3
Figure 4

We also have inspected the bifurcations with other parameters as shown in Figure 5. Except parameter $C$ and $\mu$ in Figure 5 (a) and (b) respectively, all the others remain the same as above and $\alpha$=0.04. The bifurcation diagram of the increasing of $C$ is much more like that of $\alpha$'s while when it comes to $\mu$, the composition of the diagram has different tendency with the increase of the parameter while various periods still co-exist.

Figure 5

From the diagram in Figures 3 to 5, it interestingly seems there are period-adding bifurcation sequences in the system which includes period 3. The larger the $\alpha$ is, the less the periods. If we disclose details within the lines, we could see that the line has detailed structures. Actually the lines on these figures are the coarse grained effect of huge number of points.

Figure 6

In Figure 6(a), the ascending spotted line actually is the correspondence of the horizontal line of Figure 2(a).Then in the much smaller intervals of $\alpha$ as shown in Figure 6(b), (c) and (d), the lines in Figure 2(a) are turned out to be made up by set of spots. So the periodic states in the diagram are actually the coarse grained results. We call these phenomena as coarse grained bifurcations. For any measurement in the real world, we are always restricted to certain precision. Hence these results are meaning full for us to understand the nonlinear dynamics.

B. Different attractors in the phase space

Figure 7

In the simulations presented in last subsection, we have found that with the change of parameter $\alpha$ the system will finally stabilized in the different states from the same initial condition. There are obviously three attractors in phase space as shown in Figure 2 (a). The property of the whole phase space still cannot be confirmed. To make this clear, we take another approach. We set the initial values of the number of agents as the perturbation of the existing steady attractors. In the simulation, we set the value of $N_0$ slightly changed from last steady state but keep the $\sum_{j=0}^{k} N_j = M$ constant. The result is shown in Figure 7. Over certain intervals of $\alpha$, there coexist three attractors. With the change of $\alpha$ each attractor can also become unstable and the system finally jump to other attractor. With the diminishing of $\alpha$, when period 1 become unstable, the system jump to period 2 solution. But when period 2 lose its stability, the system jumps to another attractor first, the black block around the interval (0.033-0.04) as shown in Figure 7.

As mentioned in subsection A, the orbit of the system is actually the coarse-grained periodic orbit. Other measurements like analysis of power spectrum and Lyapunov exponents should be applied to make sure its properties. We have checked the largest Lyapunov exponents of the system. The result is interesting. When we follow one orbit, for most time the exponents are zero or negatively close to zero which show the system's quasi-periodic character. While there are always a few points of the orbit seeming highly sensitive to tiny changes turning the orbit aberrant which makes the exponent positive. For example, we have shown the time series of one orbit's $\log(d_1/d_0)$ in figure 8(a). The two high points stand for the aberrance we mentioned above. Figure8(b) shows the exponents of the system for different parameter $\alpha$. In fact the positive valued points on it are caused just by those a few sensitive-to-change points. .

Figure 8

C. Other form for the return function

The form of the return functions we have taken all above is originated from the problem of the multi-agent system. It is complicated. We believe that there should be some easier or clearer forms for the function while similar dynamical outcomes are still desirable. We found that in the simulation above the function of return can be shown as quadratic curves of some parameters. Then we substitute the form of function with logistic ones

$$R(\{N_j\}) = Am(1.3-m) - C_D(\{N_j\}),$$

where $m = \sum_{j=0}^{k} N_j q_j$ and $C_D$ still takes the form $C_D = \sum_{j=1}^{k} N_j q_j (1+\alpha(1-q_j))C$ , because it is necessary for the system to give rise to the result of pattern of specialization. In the numerical simulation we set the parameter as $A$=40.0 and $C$=2.0. Similar dynamics can be expected as Figure 8 has shown.

## IV. CONCLUDING REMARKS

Organizations observed in natural and social complex systems are remarkable phenomena of pattern formation. They are usually the results of long term optimizing evolution. The evolutionary principles should play a role in the explanation of the emergent property or the organizational processes in many kinds of complex systems. Despite the diversity of time and space scales involved all these processes are governed by the same principles of competition, mutation and selection. From this point of view, we have tried to construct a dynamical model to describe the evolutionary process, and to help the exploration of the mechanism of organization formation in long-term optimizing evolution.

The Markov chain model presented in this paper gives an approach to understand the mechanism behind these innovation phenomena and demonstrates how a global structure can be generated in the evolution. The division of agents in different tasks could emerge from the long-term optimal evolution by a series of bifurcations under the mechanism of increasing returns. The dynamical behavior of the system also provides some interesting phenomena including coarse-grained periodic attractors and bifurcations.

Actually the Markov chain model is a $1d$ coupled map lattice. And the transition probability is a discontinue function. This may result in some novel dynamical phenomena. It has been reported that in an ocean internal wave model, there are chaotic solutions the period-adding sequence [13]. What we observed here is different from those results. Some detailed investigation should be done to get further understanding of the system.

## ACKNOWLEDGMENTS

The work is partially supported by the NSFC under the grant No. 70371072 and grant No. 60003018.The authors are grateful to Prof. Qiang Yuan and Dr. Yiming Ding for their helpful discussions.

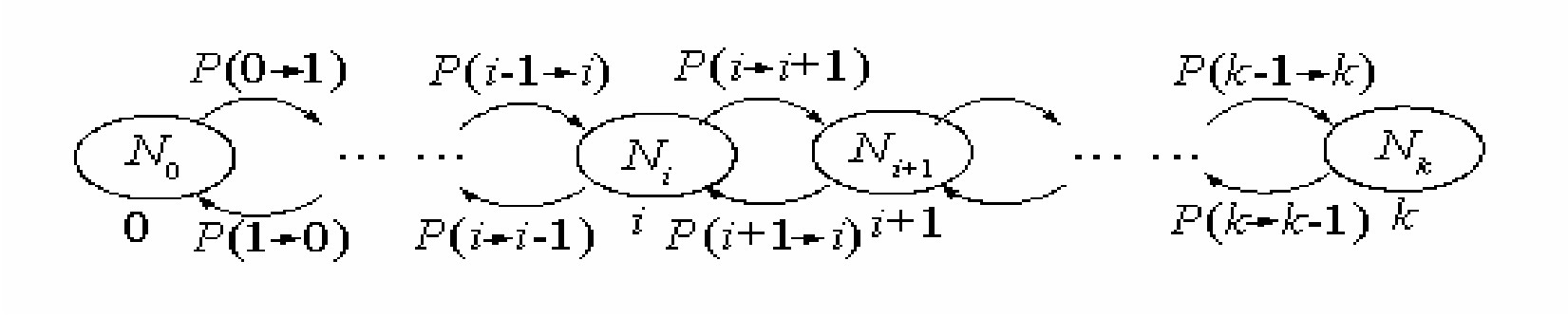

Fig.1 –Lei Chai

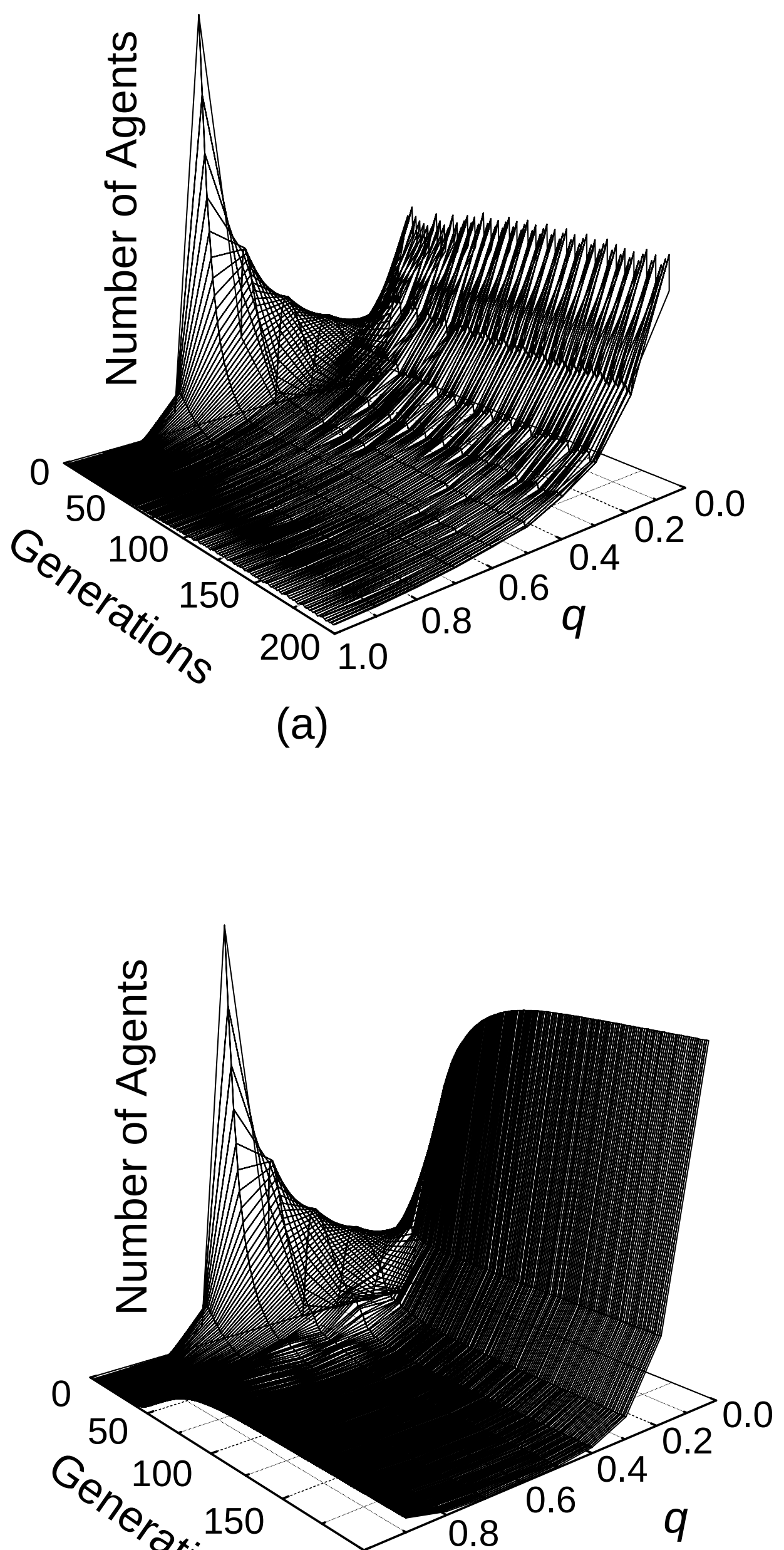

(a)

(b)

Fig.2 – Lei Chai

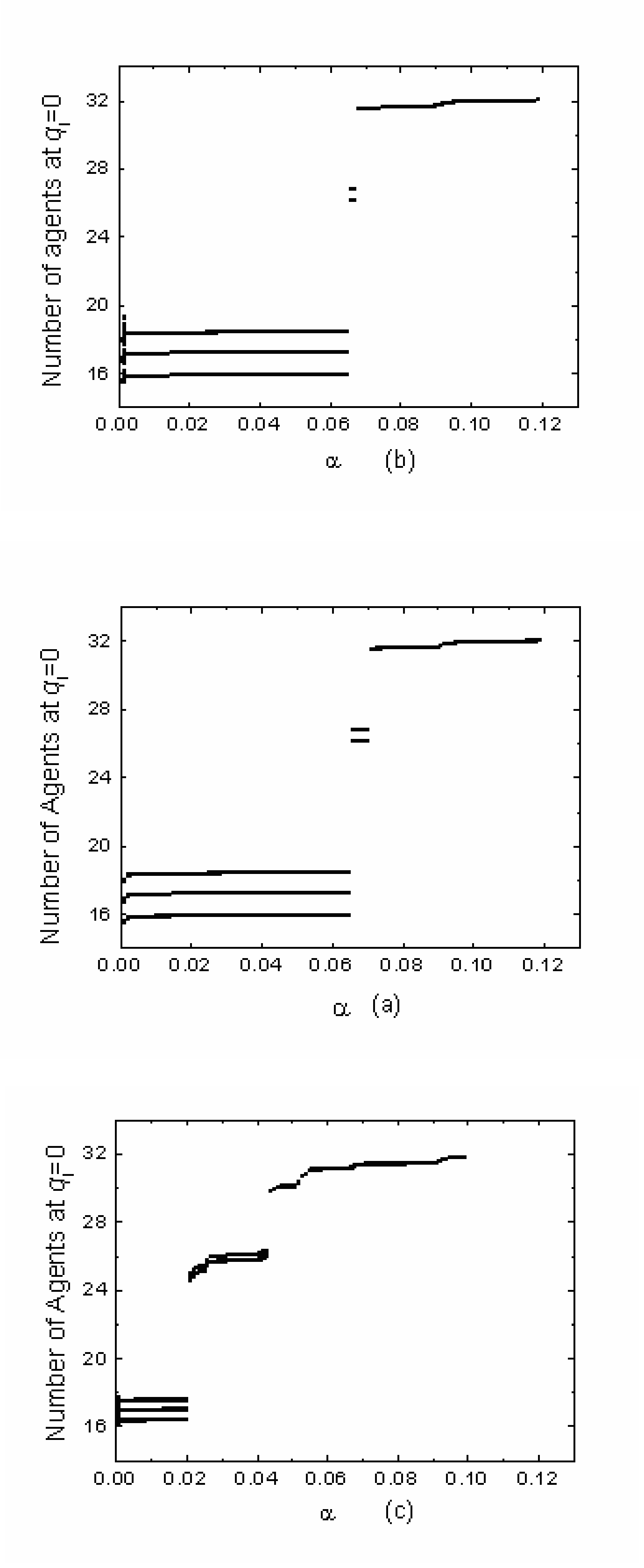

Fig.3 –Lei Chai

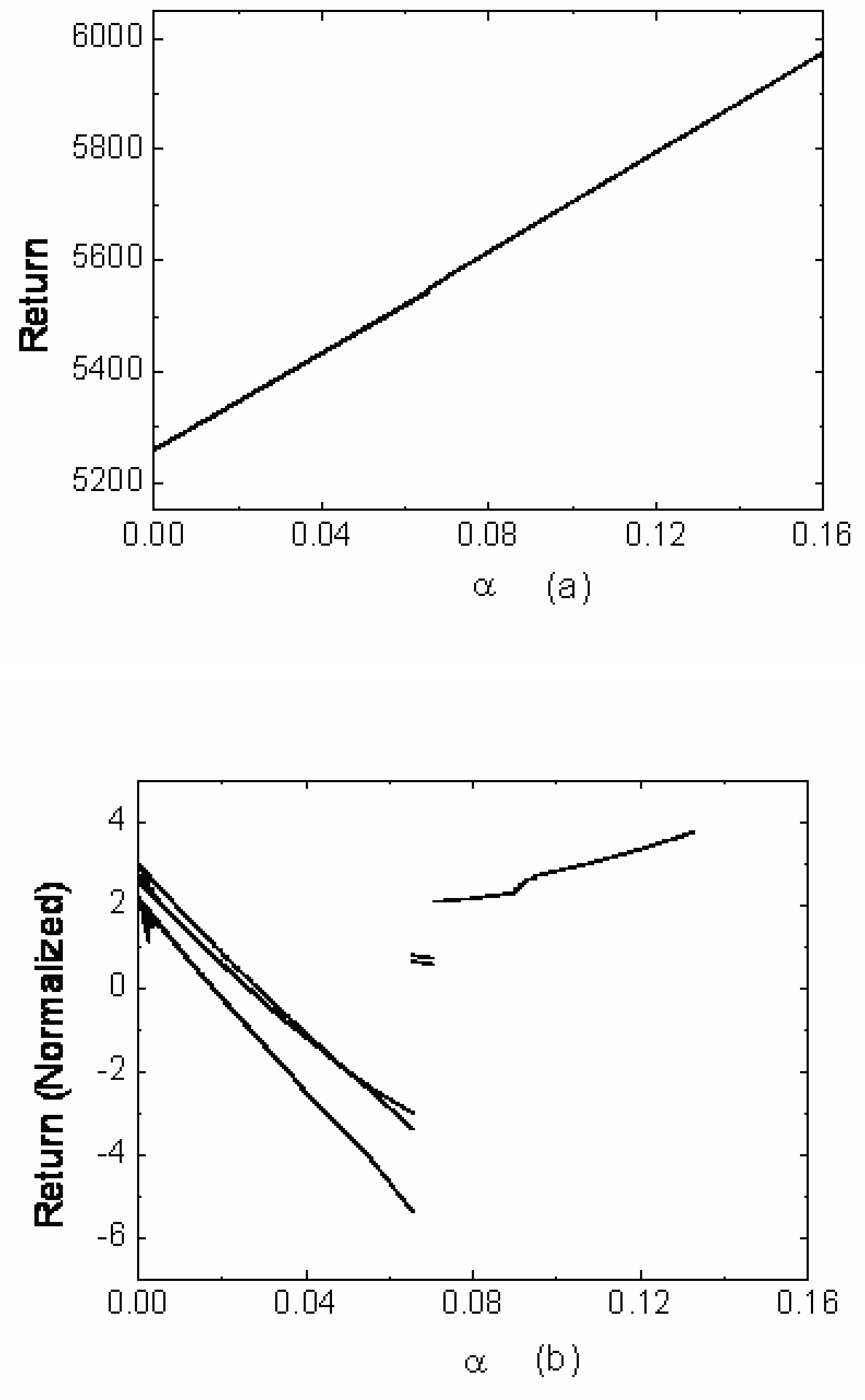

Fig.4 –Lei Chai

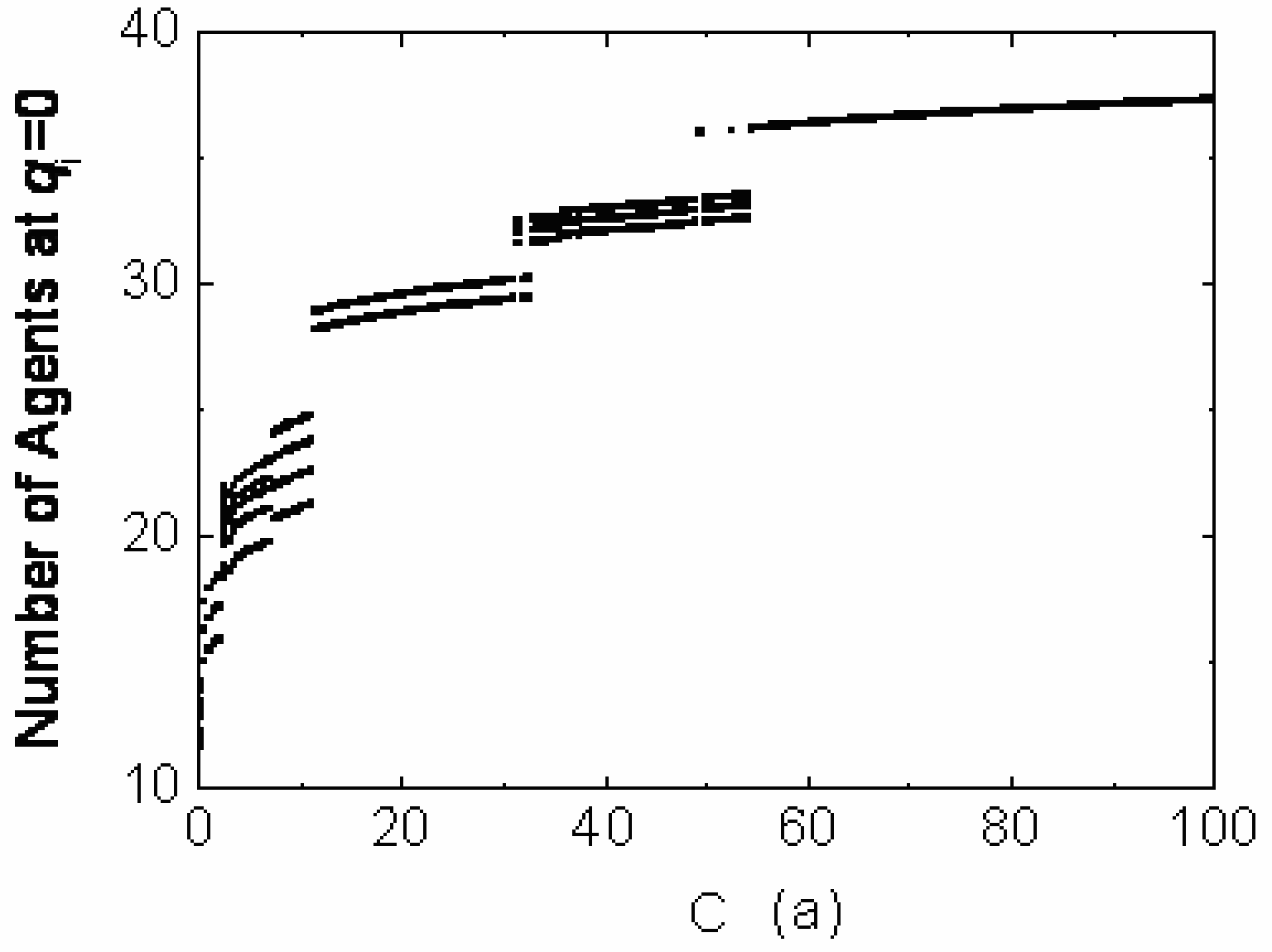

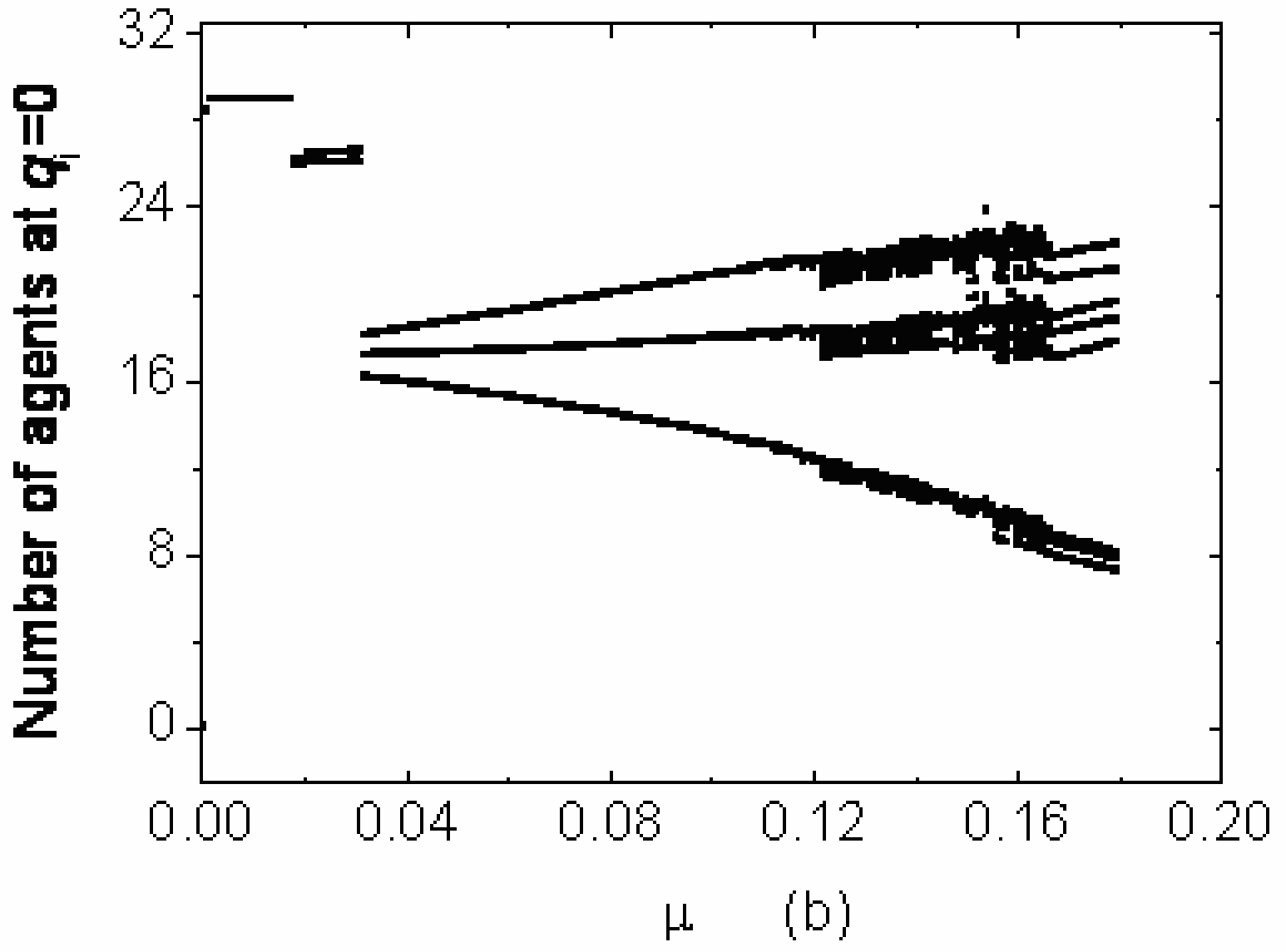

Fig.5 – Lei Chai

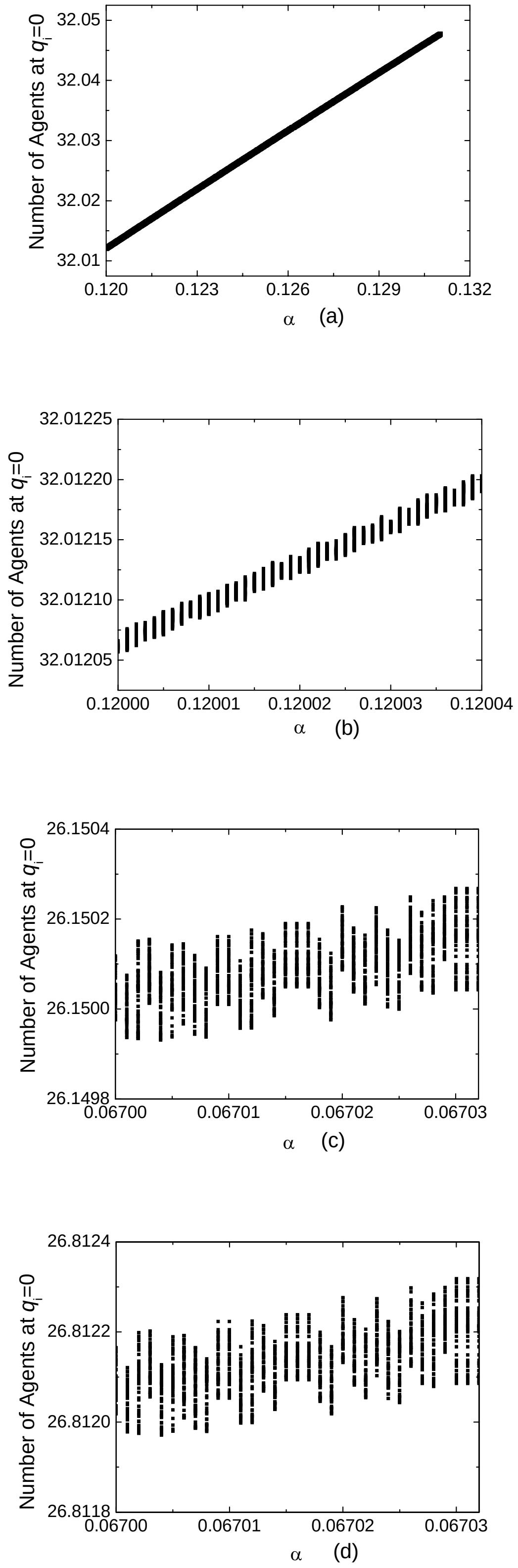

Fig.6—Lei Chai

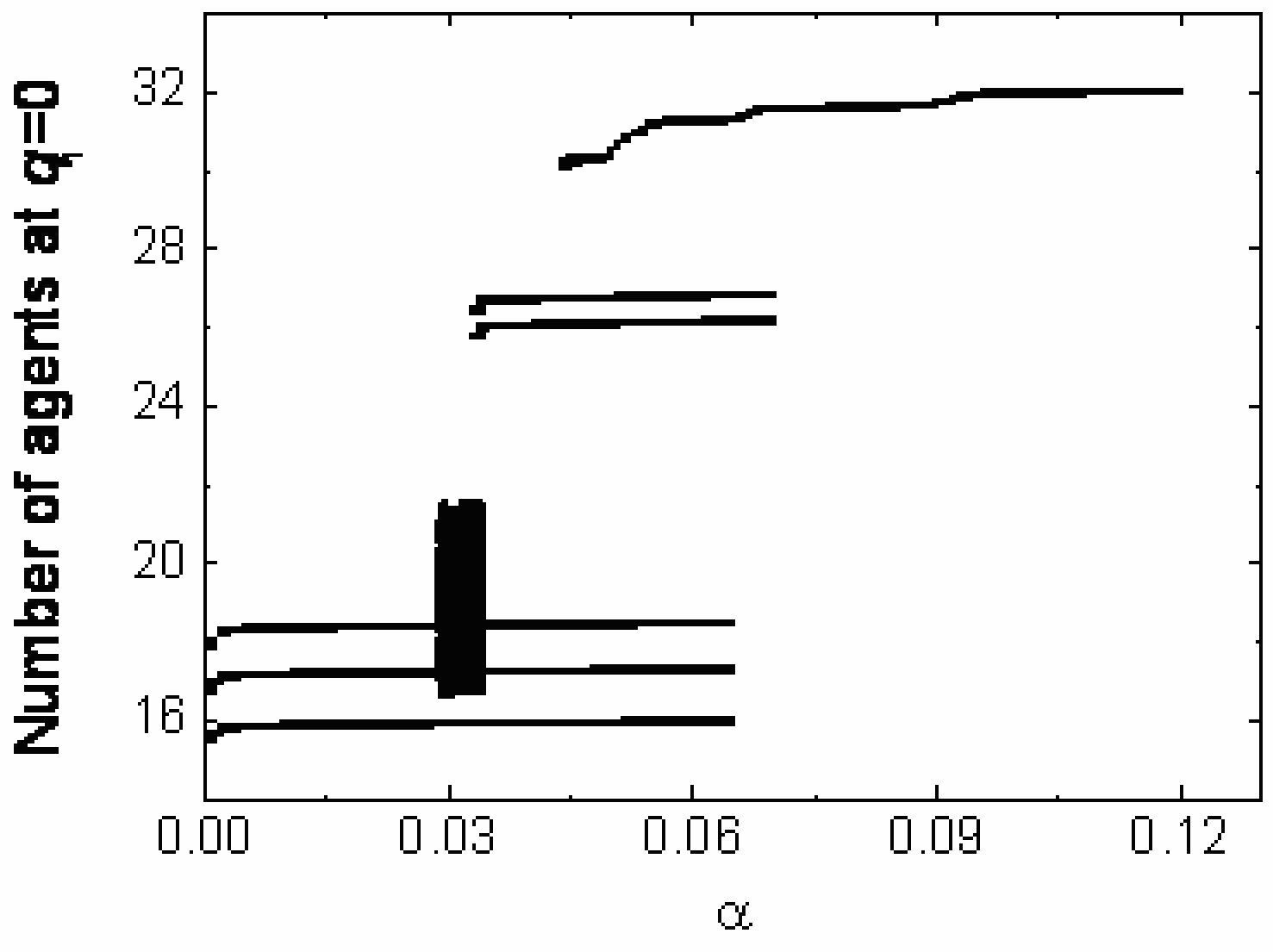

Fig.7—Lei Chai

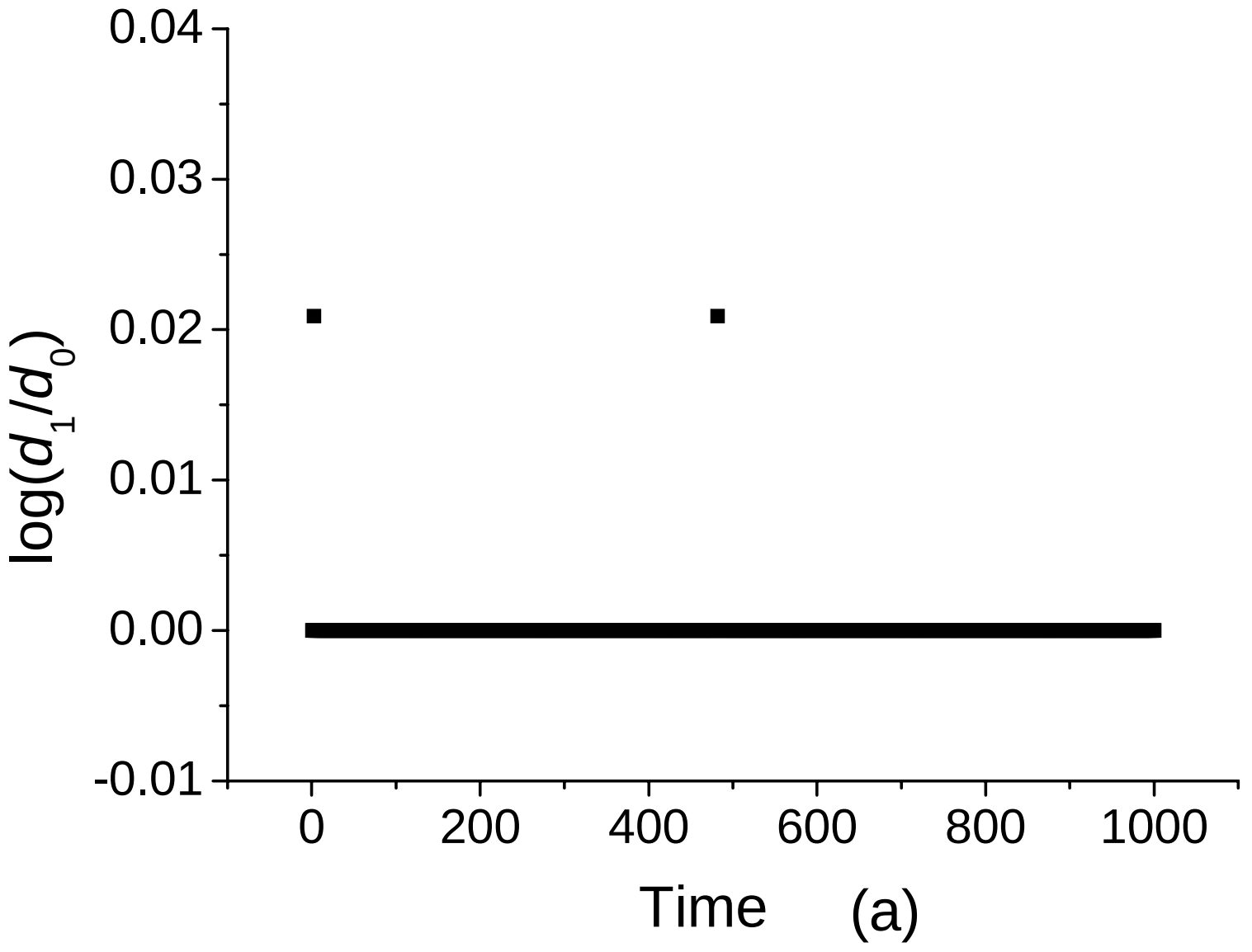

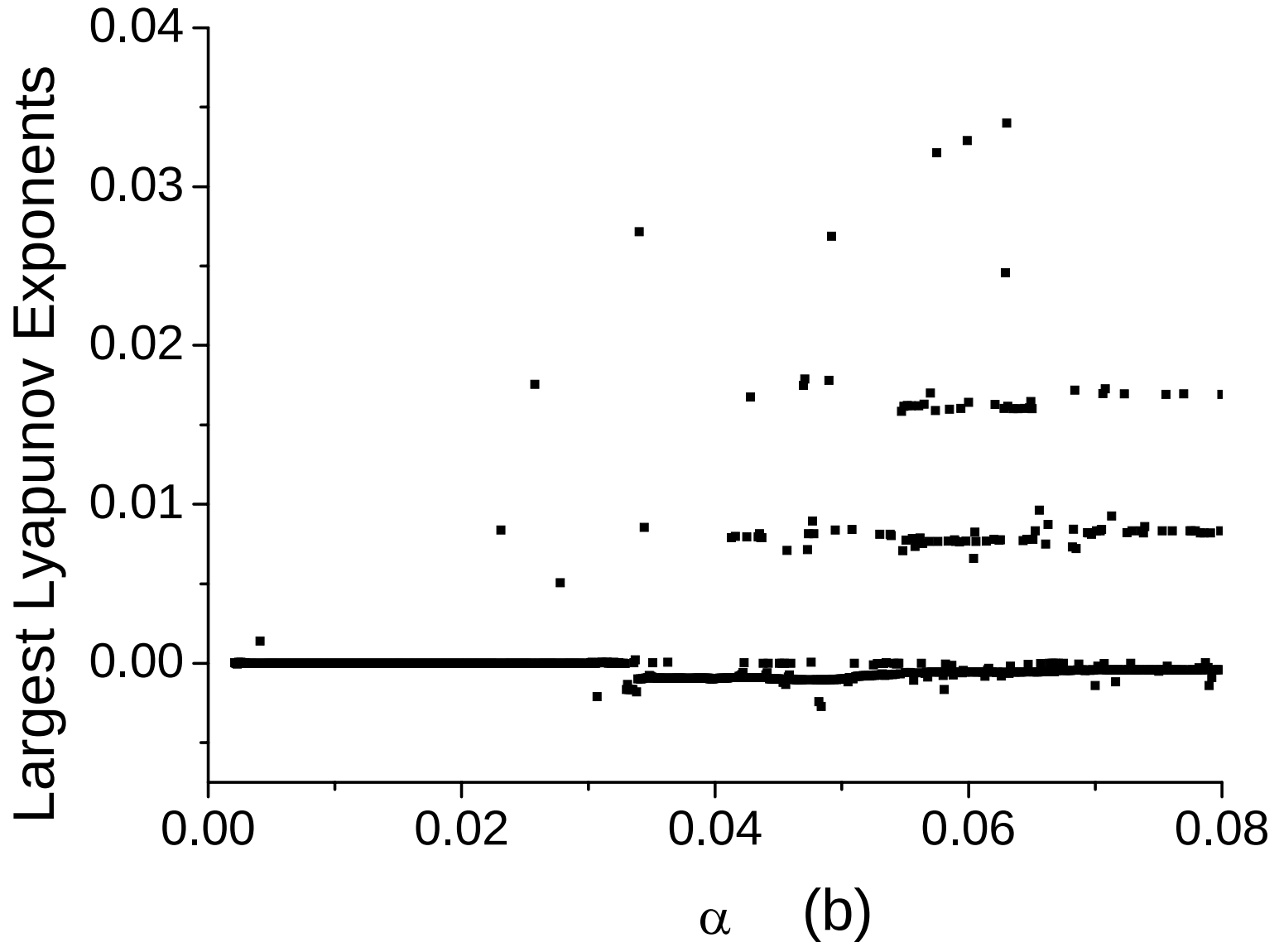

Fig.8—Lei Chai

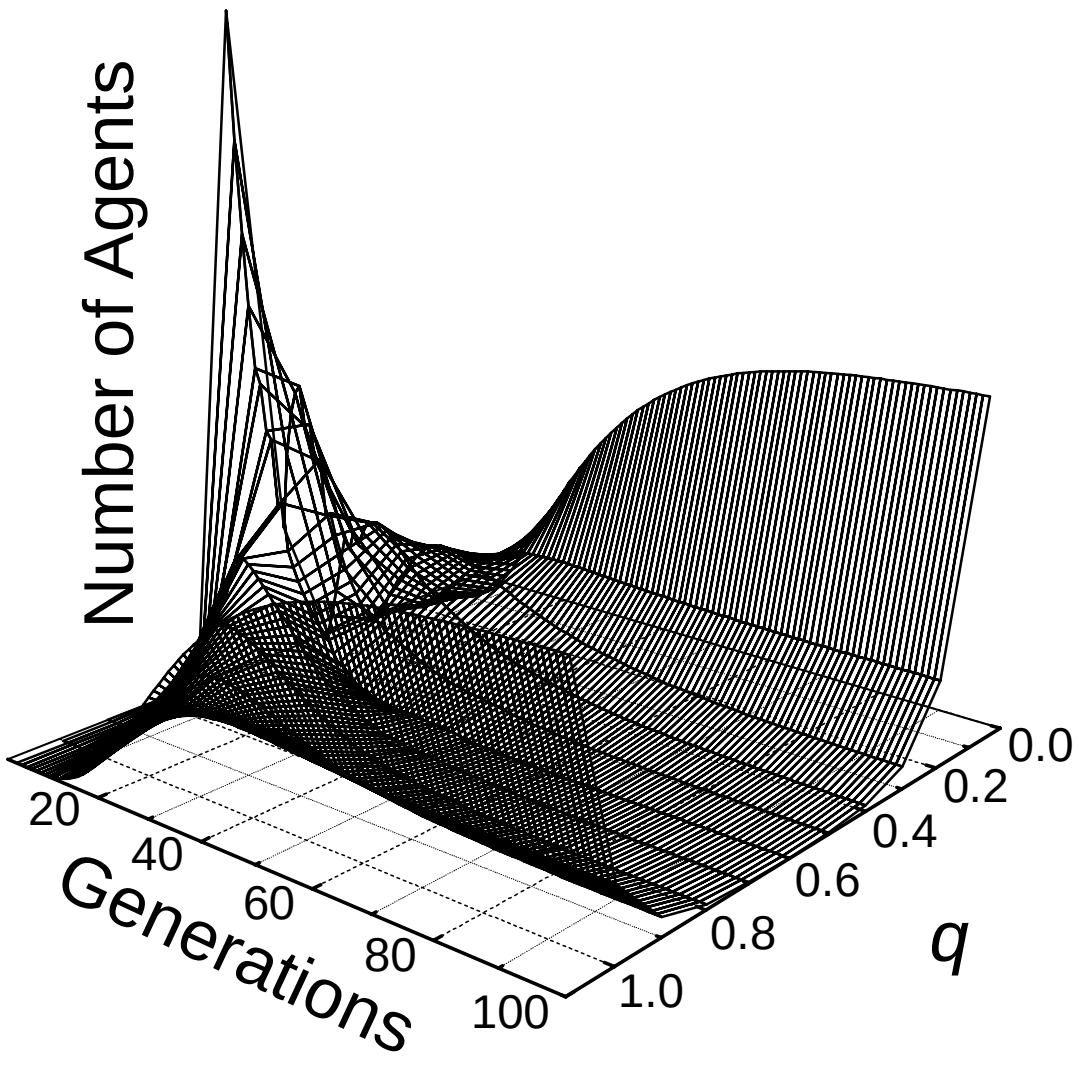

(a)

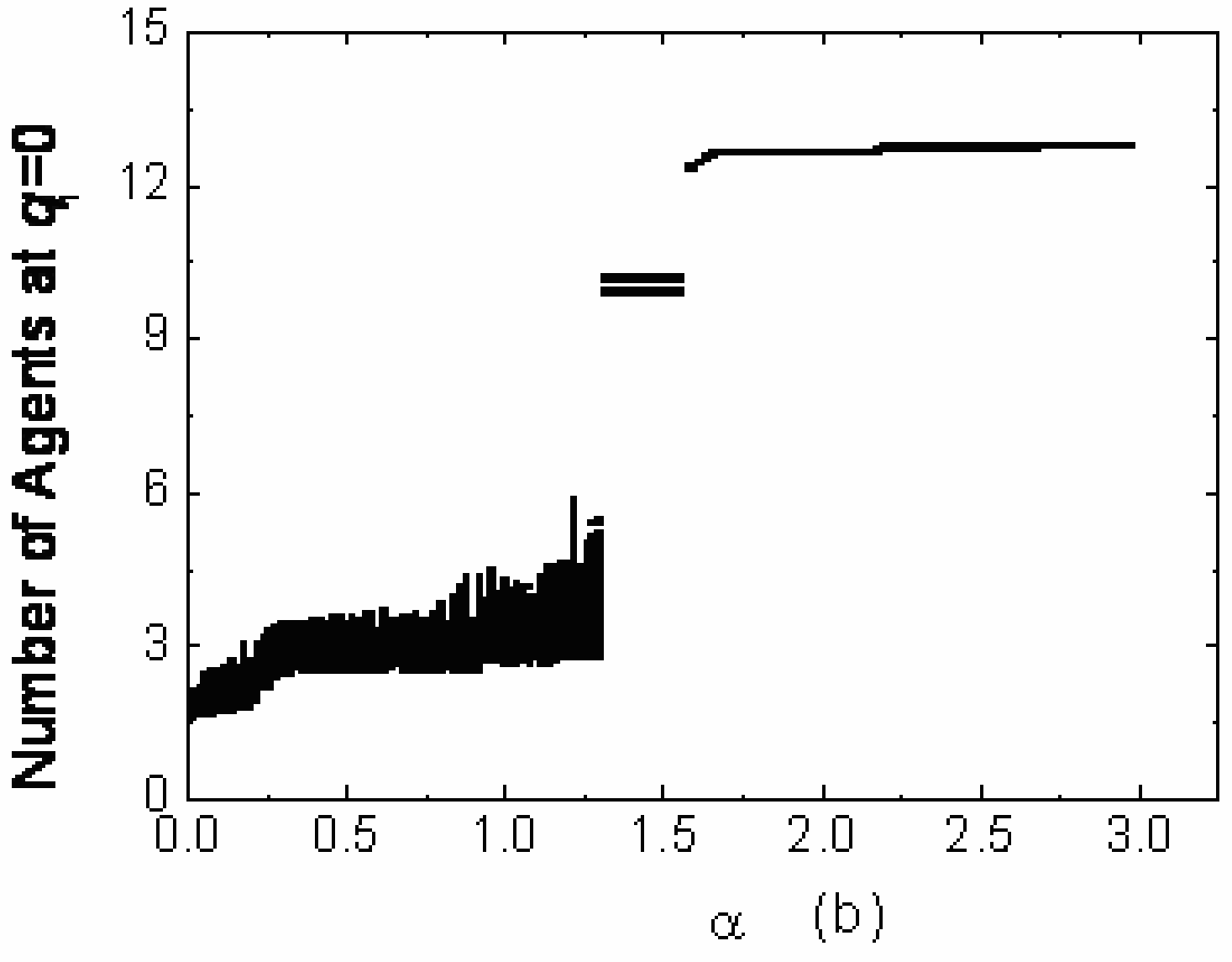

(b)

Fig.9 – Lei Chai

**Figure Captions**

FIG. 1. Markov chain process

FIG. 2. Evolution of the system. (a)α=0.04, (b) α=0.15, in which agents finally are divided into two groups in case (b).

FIG. 3. Steady state of the system (describe by the number of agents with the character $q_i$=0) with the change of parameter α. All the other parameters are given in the text. (a) all the agents initially homogeneous with $q_i$=0.5; (b) the agents randomly distributed over the interval $q_i \in [0,1]$; (c) $\mu$=0.02 and $a$=10.0 with all the agents initially homogeneous at $q_i$=0.5 and the other parameters the same as (a) and (b).

FIG. 4. Final total returns vs. parameter α, parameters and initial conditions same with Fig.1 (a). (a)Returns increasing with bigger α; (b)the total returns normalized by $W'=W-(5252.205+4500.099*\alpha)$. It has the same detailed structure as in Fig. 1 (a).

FIG. 5. Steady state of the system with the change of parameter $C$ (a) and parameter μ (b). All the other parameters given in the text and α=0.04 and all the agents initially homogeneous with $q_i$=0.5.

FIG. 6. Detailed structures of coarse-grained periodic solutions. (a) and (b) are detailed curves for period 1. (c) and (d) are two branches for period 2.

FIG. 7. Attractors in the phase space with the change of parameter α. All the other parameters the same as in Fig.1 (a). In some interval of the parameter α, the attractor coexist in the system. There are also find some other attractors.

FIG. 8. Lyapunov exponent measurement of the system. Beginning with the initial condition in the basin of attraction and after 6000 iterations the orbit is on the attractor. We select nearby point (separated by $d_0$), advance both orbits one iteration and calculate new separation $d_1$. Readjust one orbit so its separation is $d_0$ in same direction as $d_1$. After 3000 iterations repeat this we start to record log $|d_1/d_0|$. (a) is the evaluation of log $|d_1/d_0|$ over the last 1000 iterations (9001-10000). (b) is the largest Lyaponov exponents corresponding to different final steady states.

FIG. 9. Logistic form of benefit functions. (a), evolution of the system; (b) the change of final steady state vs. parameter α.